\documentclass[preprint]{elsarticle}
\usepackage{graphicx}
\usepackage[english]{babel}
\usepackage[utf8]{inputenc}
\usepackage{graphicx}
\usepackage{amsmath}
\usepackage{amsfonts}
\usepackage{amssymb}
\usepackage[colorlinks]{hyperref}
\usepackage[thinlines]{easytable}

\date{}

\begin{document}

\begin{frontmatter}
\title{Information reduction in a reverberatory neuronal network through convergence to complex oscillatory firing patterns}
\author[1]{A. Vidybida\corref{cor1}}
\ead{vidybida@bitp.kiev.ua}
\ead[url]{http://vidybida.kiev.ua/}
\author[2]{O. Shchur}
\ead{olha.schur@gmail.com}
\cortext[cor1]{Corresponding author}
\address[1]{Bogolyubov Institute for Theoretical Physics,
Metrologichna str., 14-B, Kyiv 03680, Ukraine}
\address[2]{Taras Shevchenko National University of Kyiv,
Volodymyrska str., 60, Kyiv 01033, Ukraine}

\begin{abstract}
We study dynamics of a reverberating neural net by means of computer simulation.
The net, which is composed of 9 leaky integrate-and-fire (LIF)
 neurons arranged in a square lattice, is fully connected
with interneuronal communication delay proportional to the corresponding distance.
The network is initially stimulated with different stimuli and then goes freely.
For each stimulus, in the course of free evolution, activity either dies out completely
or the network converges to a periodic trajectory, which may be different for different stimuli.
The latter is observed for a set of 285290 initial stimuli which constitutes 83\% of all stimuli applied.
By applying each stimulus from the set, we found 102 different periodic end-states. 
By analyzing the trajectories, we conclude that neuronal firing is the necessary
prerequisite for merging different trajectories into a single one, which eventually
transforms into a periodic regime.
Observed phenomena of self-organization in the 
time domain are discussed as a possible model for processes 
taking place during perception. The repetitive firing in the periodic regimes
could underpin memory formation.
\end{abstract}

\end{frontmatter}

\section[Introduction]{Introduction}

Recordings from the inferior temporal cortex (IT) have revealed cells with unusual properties
 compared with cells in the primary visual cortex, \cite{Gross1984,Rust2010,DiCarlo2012}. 
Firstly, those cells have extremely extended
receptive fields. Secondly, those cells are selective for complex features/objects,
such as complex shapes. Thirdly, this selectivity is invariant with respect to transformations,
such as changing of object's position in space, its orientation, size, color, texture, contrast,
location on the retina,\cite{Tang2004,Durbin1990}. 
Furthermore, 
the response of those cells is stable with changing context
and the presence of distractors (tolerance to noise).
It looks like primary visual signals are stripped of redundant information 
during processing in the visual pathways \cite{Barlow1961,Barlow2002,Durbin1990}.
It is interesting that similar redundancy stripping is characteristic of
sensory pathways of other modalities, such as hearing \cite{Creutzfeldt1980} and
 olfaction \cite{Wehr1996,Duchamp-Viret1997} as well as in the primary visual areas,
 see review in \cite{Felleman1991}.

At the sensory periphery, information about external world contains all data
which primary sensory units, such as cones and rods, or olfactory receptor neurons are able 
to sense and transmit further. In a sense, representation at the sensory periphery may be
considered as continuous, \cite{Knig2006}. On the other hand, in farther areas of the brain,
 like IT, or multimodal
areas, where recognition and perception take place, representation of the real world
is conceptual and as such has a binary, discrete nature. This process of transformation of sensory
inputs from continuous, redundant representation to binary conceptual representation 
accompanied with and enabled by the redundancy reduction is sometimes called
as information condensation, \cite{Knig2006}. Specifically in vision,
the term ``dimension reduction'' is as well used, \cite{Durbin1990}.

We now propose a question: What might be the physical mechanism of the information condensation? 
The visual perception is fast: 50 ms period of image presentation is enough
to form corresponding perception state 100-150 ms later. It is hard to expect
structural changes in the neuronal network during that short period.
If we believe together with H. Barlow \cite{Barlow1972} and F. Crick \cite{Crick1995a}
that  ``the activities of neurons, quite simply, {\em are} thought processes'',
then we must expect that it is the spatiotemporal pattern of 
neuronal activity which changes during 
perception process and eventually forms a state of perception of the object
presented to sensory periphery. 

A possible mechanism of a proper change
was proposed by D. MacKay as self-organization in the time domain \cite{MacKay1962}.
Due to self-organization, the spatiotemporal dynamics of neural activity in a
pulse-propagating network is able to develop the same regime in response
to several different initial stimuli
which could be treated as redundancy reduction. This mechanism was checked 
in \cite{Vidybida2011} for a network composed of 5 binding neurons. It was found
that that simple network can have several hundreds of different periodic
regimes, each being the result of stimulation by any stimulus
from the corresponding set. Obtaining the same final regime in 
response to different stimuli means that corresponding dynamical trajectories
converge and finally merge in the time course and this might be that process which makes
definite cells in the IT insensitive to certain differences in 
the primary visual stimuli.

In the theory of mechanical dynamical systems, convergence of trajectories is a
well known phenomenon that is a consequence of mechanical/viscous friction.
Systems of this type can be described by differential equations amenable to
standard stability analysis. In a pulse propagating neuronal network, 
neuronal leakage is the analog of friction in a mechanical system.
This is because due to the leakage the electrical energy is dissipated to heat
similarly as does mechanical energy due to mechanical friction.
 But the dynamics of spiking neurons is not smooth, which makes these systems 
 unsuitable for standard stability analysis. 
 This may account for the relative paucity of studies of these systems.
Among those studies, a network of spiking
neurons considered in \cite{Herz1995,Hopfield1995a} has zero interneuronal communication delays.
The net is
composed of neurons without leakage and is in the presence of constant and permanent external
excitatory stimulation to each neuron. In \cite{Gerstner1996a}, a similar system is analyzed,
 but finite communication delays are taken into account. In spite of
visible limitations, like usage of a single stimulus, namely a constant
excitatory input to each unit, interesting conclusions are made 
in \cite{Herz1995,Hopfield1995a,Gerstner1996a}. Among those is the existence 
of many periodic regimes, which are reached after single triggering of each neuron. 

The purpose of this investigation is to figure out possible elementary physical processes
and events underpinning convergence of different dynamical trajectories,
evoked by different initial stimulation
 of a reverberating neural net,
with eventual merging them into a single one and further engaging in a periodic regime.

In the context of perception, we pay a special attention to periodic regimes for the 
following reason. 
An effective perceptual operation involves the ability to report out what has been perceived,
 see e.g. \cite{Barlow2002}. This is in contrast to reception, where 
the fastest reaction might bring the best result, e.g. \cite{Camhi1978}.
The ability to report supposes that perceived object is held in a short-term memory of some sort.
This is in concordance with the finding of several types of memory in the IT, see
\cite{Miyashita1993}. The short-term memory through long-term potentiation
requires repetitive stimulation of the same synapses,\cite{ Andersen2003},
and this can be well achieved during periodic dynamics in the neuronal network.

Indirect experimental evidence that some neuronal assemblies can be engaged into a periodic-like
regime is vast, \cite{Baar2004,Traub1999,Buzsaki2006}. 
Among those are stimulus-evoked oscillation-type activities, see 
\cite{Eckhorn1988,Eckhorn1990,Gray1989,Engel1991a}. Activities of single neurons are 
observed to be phase-locked with each other, but not with the stimulus presentation moment.
 This suggests that those
activities have a self-organized, emergent nature. Similar suggestions might be derived from
observation of so-called unitary events in the motor cortex during preparation and execution
of movements, see \cite{Riehle1997,Grn2010,Grn2009}.

\section{Methods}

\subsection{Neural network}
\begin{figure}[h]
\begin{center}
\includegraphics[width=0.4\textwidth]{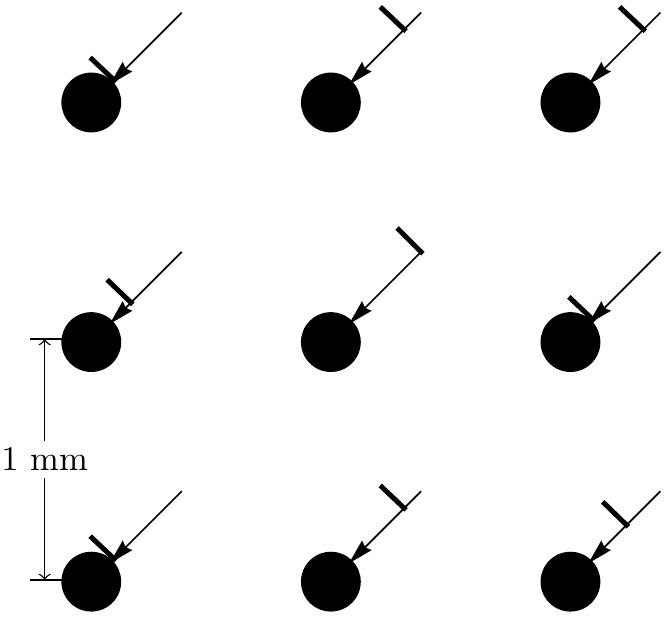}
\end{center}
\caption{\label{net}Network used in simulations. The network is fully
  connected (the internal connections are not shown).
   The interneuronal communication speed $v=1$ m/s. The communication delay
   between neurons with the positions $(i,j)$ and $(k,l)$,
is calculated based on the Euclidean distance:
$d_{(i,j),(k,l)}=1\,mm\cdot\sqrt{(i-k)^2+(j-l)^2}/v$.
Each input impulse got from internal connection rises
  the LIF's excitation level by 2.71 mV.
The internal synaptic weights, $W_{(i,j),(k,l)}=1$ for all $i,j,k,l =0,1,2$,
$(i,j)\ne(k,l)$.
 Impulse received from an external connection (shown by arrows) rises
excitation just above the firing threshold.
This corresponds to the input synaptic weights $W^{input}_{(i,j)}\ge 7.39$.
}
\end{figure}

\begin{figure}[h]
  \begin{center}
        \includegraphics[width=0.45\textwidth]{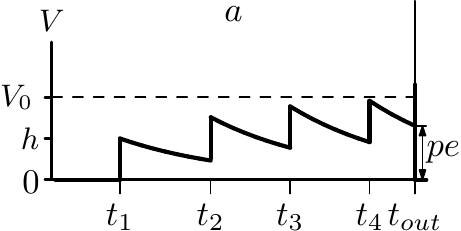}
        \hfill
        \includegraphics[width=0.45\textwidth]{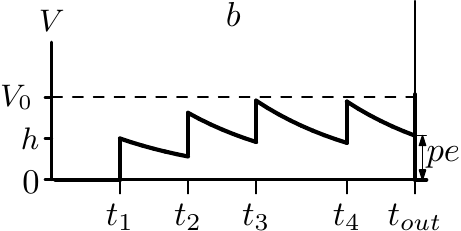}
\end{center}
\caption{\label{SiCon}Example of two different stimuli into a single LIF neuron,
  which evoke identical outputs. 
 Each neuron has a firing
  threshold $V_0=20$ mV,  The relaxation time of any LIF neuron
  $\tau=20$ ms.  
  The triggering input spike is delivered at time $t_{out}$.
  Both stimuli evoke the same output (a spike at the same moment) 
   in spite of having a different temporal position of input spikes \#2 -- \#4.
The different timing results in different pre-triggering excitation, $pe$ in the
figures. The $pe$ difference is cancelled after triggering.
}
\end{figure}

\begin{figure}[h]
\begin{center}
\includegraphics[width=0.5\textwidth]{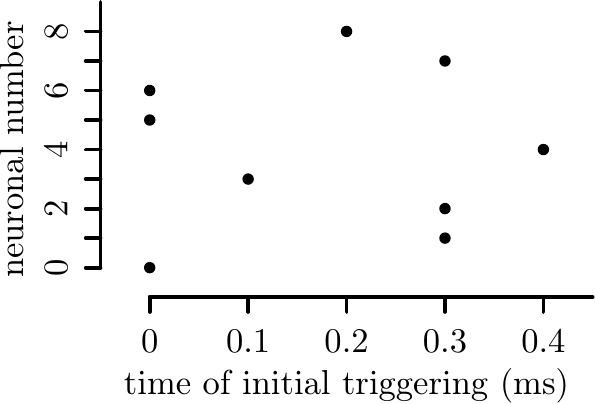}
\end{center}
\caption{\label{StiExample} Example of a single stimulus to start a reverberating
  dynamics. Moments of initial triggering shown here exactly correspond to positions
  of bars in Fig. \ref{net}. Each dot specifies the triggering time of corresponding
  neuron (vertical axis) due to the stimulus. Neuron \#0 is always triggered 
  at the moment when a stimulus starts.}
\end{figure}

For performing simulations we use a network of nine excitatory leaky
integrate-and-fire (LIF) neurons organized into a square lattice of definite
geometric size, Fig.\ref{net}. The interneuronal communication delays are calculated
based on the interneuronal distance and the communication speed.
Each neuron is characterized by the same firing
threshold, $V_0$, the relaxation time, $\tau$ and the unitary EPSP height, $h$, see legend
to Fig.\ref{net},\ref{SiCon} for the numerical values.
The network is fully connected and each neuron has additional external excitatory
input used for delivering initial stimulation necessary to start reverberating
dynamics. Example of a single stimulus for the whole network
is given in Figs. \ref{net},\ref{StiExample}.

The network states during evolution are generated with the time step $dt=0.1$ ms.
Interneuronal communication delays, which do not equal to a whole number of $dt$,
are rounded, e.g. delay through the diagonal is taken 2.8 ms instead of
$10\sqrt{8}\,dt$. The LIF states are expressed in terms of integers as well, see
Sec. \ref{intLIF} below. This allows us to run evolution by means of integer
arithmetics and to figure out exact periodic regimes of  the net.

\subsubsection{Simulating LIF neuron with integers}\label{intLIF}
Dynamics of a spiking neural network can be unstable. This is because in some dynamical
states infinitely small delay in arriving moment of input triggering spike
may prevent neuron from achieving
the firing threshold at that moment. Then the neuronal triggering will be
initiated by the next or further input spike. As a result, the triggering time will
change considerably and this change can be amplified during further evolution.
Instability requires that, in the search of a periodic regime, states are reproduced exactly
after the regime's period.
On the other hand, exact equality cannot be stated in computation with floating-point numbers, see \cite{Goldberg1991} for discussion.

In order to overcome this difficulty, an approximation method has been
developed, \cite{Vidybida}.
In this method, the whole set of LIF states, namely the interval of real numbers $[0;V_0]$,
is approximated by a finite set of discrete values unevenly distributed in
the $[0;V_0]$. Each discrete value is labeled by a pair of integers $(n,i)$.
Either receiving of input impulse or evolving during simulation time step $dt$
results in changing the pair. This approximates evolution of LIF states
expressed in terms of floating point numbers by the evolution of states expressed in terms of 
integer labels
and makes them amenable for exact comparison in
computer. The two models are denoted as fpLIF and intLIF.
The quality of such approximation is checked by generating triggering moments
of both fpLIF and intLIF under fairly long random inputs.
It was established that with suitable choice of approximation
parameters the sequences of firing moments of both models match with each other
 exactly if expressed in terms of a
whole number of $dt$. See \cite{Vidybida} 
for details.

\subsection{Data acquisition}
The purpose of running the simulation was to characterize the input-out behavior of the network, i.e. to enumerate all periodic regimes to which the network evolves given an 
initial stimulus taken from a definite set  $\mathrm S$.
We mean here that the network can settle into some periodic firing pattern that was 
not necessarily related in any simple way to the stimulus. This periodic firing pattern
is attained due to self-organization in the time domain. Many different initial stimuli
may result in the same periodic pattern.
The set of stimuli has elements $\{t_0=0,t_1,\dots,t_8\}$, where $t_i$
denotes the moment (in $dt$ units) of triggering of neuron \#$i$ due to the stimulus,
see Fig. \ref{StiExample}.
Each $t_i$ can take any value from $\{0,1,2,3,4\}$ except $t_0$, which is
always 0. We leave initial firing moment for the
	neuron \#0 invariable in order to eliminate degeneration due to 
	the uniform shift of initial firing moments for all neurons.
	  Thus, the first stimulus in the set $\mathrm S$ 
is $\{0,0,0,0,0,0,0,0,0\}$ and the last
one is $\{0,4,4,4,4,4,4,4,4\}$ with total number of stimuli $5^8=390\,625$.
For each individual stimulus, we consider the moment of arrival of last
input triggering spike from that stimulus as $t=0$ for the subsequent net's evolution.

After applying a stimulus from $\mathrm S$, the net's evolution had been running
 with time step $dt$ without external stimulation
until a state identical to already observed in this run occurred. Since the dynamics
is deterministic, the occurrence means that a periodic regime has been achieved.
The entire sequence of states covering the periodic regime is enclosed between the first
and the second occurrence of  the two identical states, one of them included.
The result of each run was stored in two Berkeley Database tables.
The first table, IniNum.db, has the initial stimulus as its first column.
The second column is composed as a pair 
of corresponding periodic regime serial number and relaxation time required to entrain on 
that regime after the end of the 
initial stimulation. For some stimuli dynamic decays to the
dead state with no triggering. This state was numbered as 0.
The second table, NumPesta.db, has a representative of the periodic regime 
as its first column. Actually, this was exactly the repeated state mentioned above.
The second column is the regime's serial number. Kicked with two different stimuli, the net
may entrain onto the same regime in two different points of the cycle, see Fig. \ref{route}(C).
In order not to count
these cases as two different regimes, we have performed an adequate checking.

For observing the convergence progress during approaching the same cycle,
we have generated the whole set of trajectories corresponding to that cycle
and then have analyzed them with additionally developed programs.

\section{Results}

\subsection{Periodic regimes and their domains}

\begin{figure}
\includegraphics[width=0.55\textwidth,angle=-90]{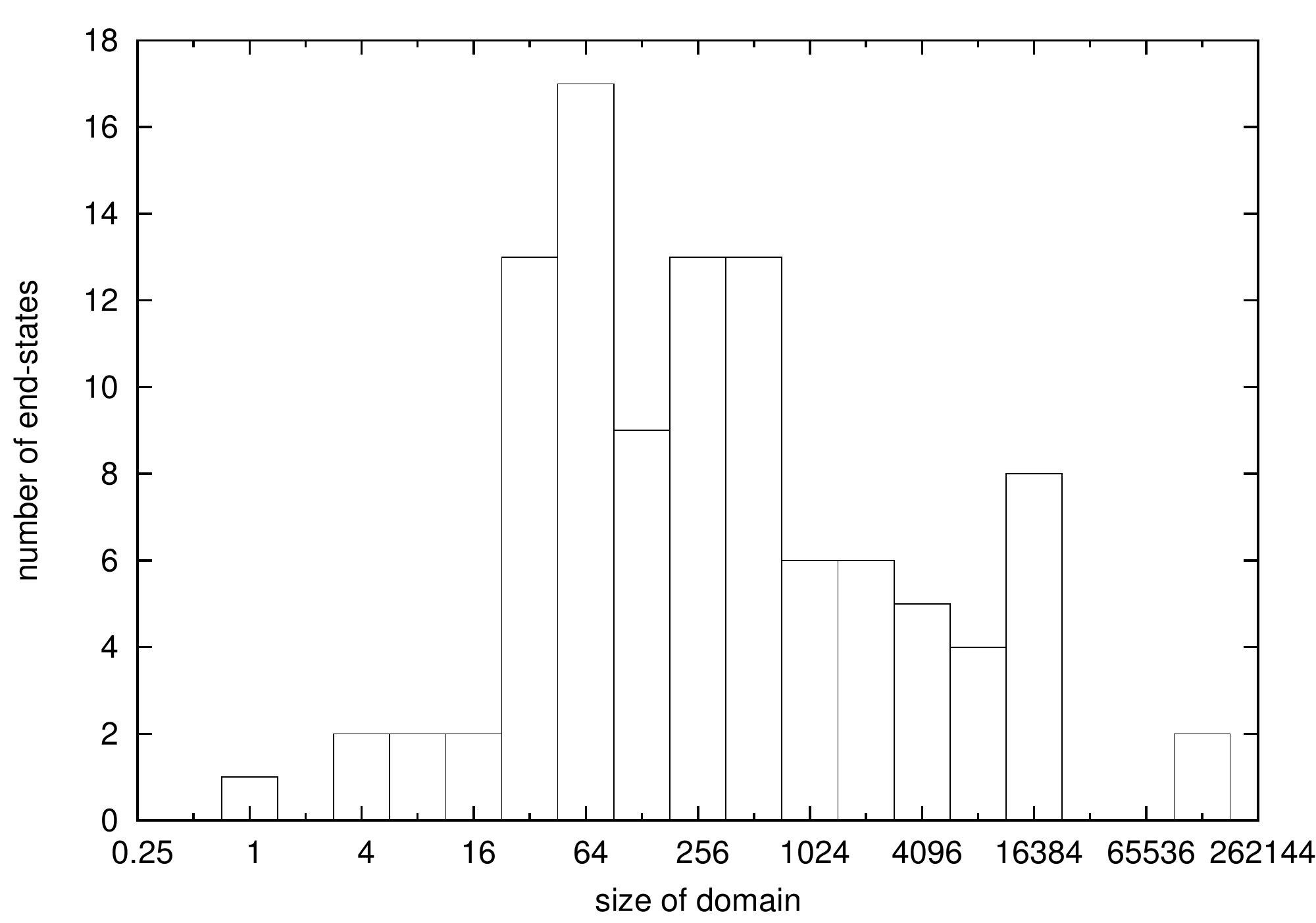}
  \caption{\label{histo}Histogram of sizes of attractor domain.
  This is for 285290 stimuli giving rise to one of the 102 periodic end-states. 
  Another 105335 stimuli giving rise to the dead end-state are not presented here.}
\end{figure}

After trying all stimuli from the set $\mathrm S$, we have obtained 102 different
periodic regimes. This splits $\mathrm S$ into 102 + 1 disjoint subsets
corresponding to different periodic regimes\footnote{In paper \cite{Vidybida2011}
  for a network of 5 neurons it was found much more, namely 485
  different periodic regimes. This is because in \cite{Vidybida2011} moments
  of initial triggering could take 100 different values which gives for the size
  of set $\mathrm S$ as much as 100\,000\,000 stimuli instead of 390\,625 as in this work.
  Extending the set $\mathrm S$ might bring about more periodic regimes
  for our network. This is expected to check in future work.}. ``1'' stays for the dead regime.
The histogram of sizes of the subsets is given in Fig.\ref{histo}.

In the theory of dynamical systems, the term ``attractor basin'' is used.
The basin is composed of all points in the phase space of a system, which
approach corresponding attractor due to dynamic. Here we consider not phase points
(each phase point is composed of states of all individual neurons and axons)
but points in the set of stimuli (each point is composed of the moments of 
delivering initial triggering stimulus for each neuron). Therefore, we use 
for the above subsets the term
``attractor domain'' instead of ``attractor basin''.

Periods of the regimes found are given in Table \ref{peri}.

\begin{table}
\begin{TAB}(@,1cm,1cm)[5pt]{|c|c|c|c|c|c|c|c|c|}{|c|c|}
    period (ms)      & 3.0& 3.2& 3.4& 3.6& 4.0& 6.8& 7.2& 10.4\\
     number of regimes & 5& 20& 11& 29& 19&  6&  6&  6\\
\end{TAB}
\caption{\label{peri}  Number of regimes with a given period.}
\end{table}

For a given periodic regime, each of the 9 neurons fires the same number of spikes during the whole cycle.
This number depends on the period's length as shown in Table \ref{spikes}.

\begin{table}
\begin{TAB}(@,1cm,1cm)[5pt]{|c|c|c|c|}{|c|c|}
    spikes per period       & 1& 2& 3\\
    period duration (ms) & 3.0, 3.2, 3.4, 3.6, 4.0& 6.8, 7.2 &10.4\\
\end{TAB}
\caption{\label{spikes}Number of spikes per period.}
\end{table}

\subsection{Routes to periodic regime}

\begin{figure}

\includegraphics[width=0.55\textwidth,angle=-90]{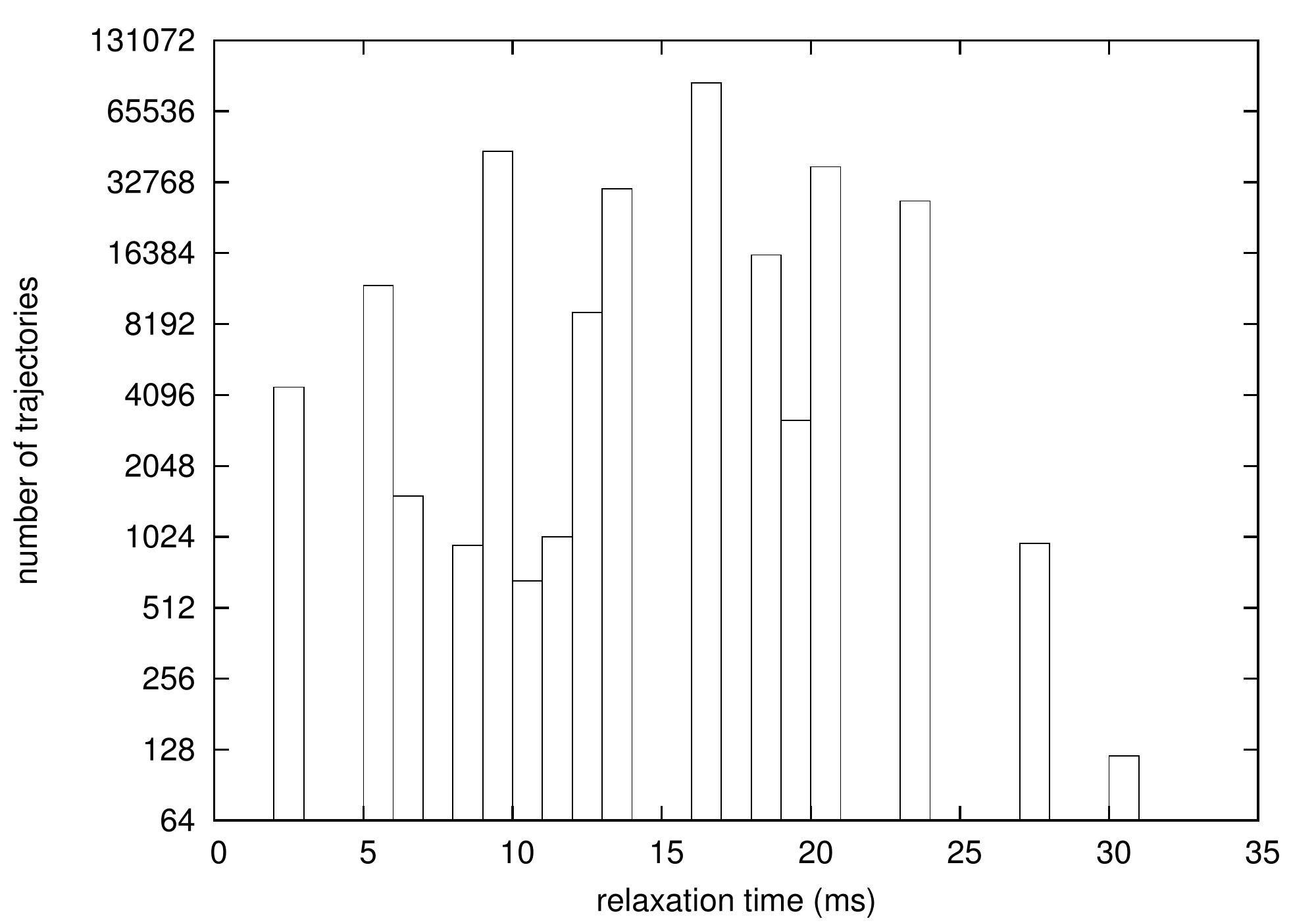}
  \caption{\label{Rhisto}Histogram of relaxation times required for attaining a periodic regime.}
\end{figure}

Before the dynamics
enter into the periodic regime, each neuron has been triggered 1 to 10 times. 
The relaxation times required to enter the
periodic regimes are shown in Fig. \ref{Rhisto}. In some cases, all trajectories
belonging to a single periodic regime have the same duration, 
in other cases the trajectories may have several different duration.

\begin{figure}
  \unitlength=1mm
  \begin{picture}(0,3)(10,0)
  \put(30,0){A}
    \put(70,0){B}
      \put(110,0){C}
  \end{picture}\\
    \includegraphics[width=0.30\textwidth]{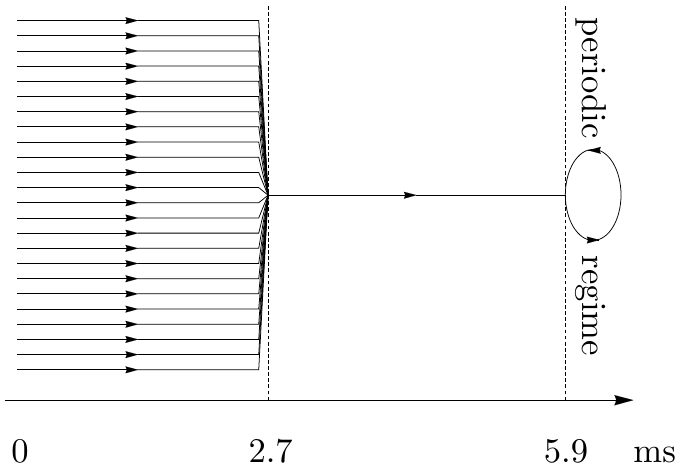}
    \hfill
      \includegraphics[width=0.30\textwidth]{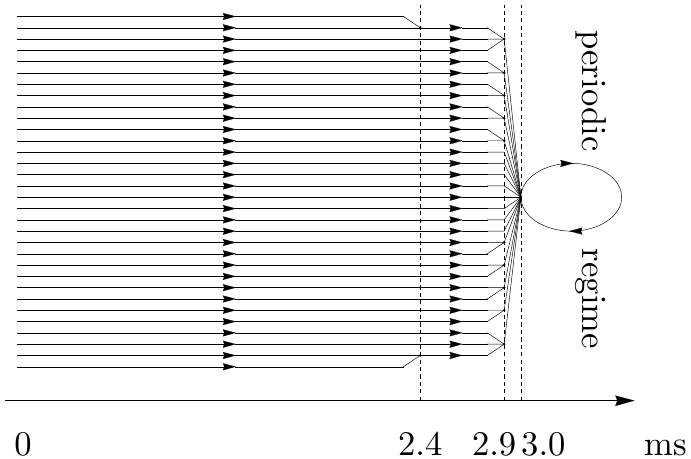}
  \hfill
  \includegraphics[width=0.30\textwidth]{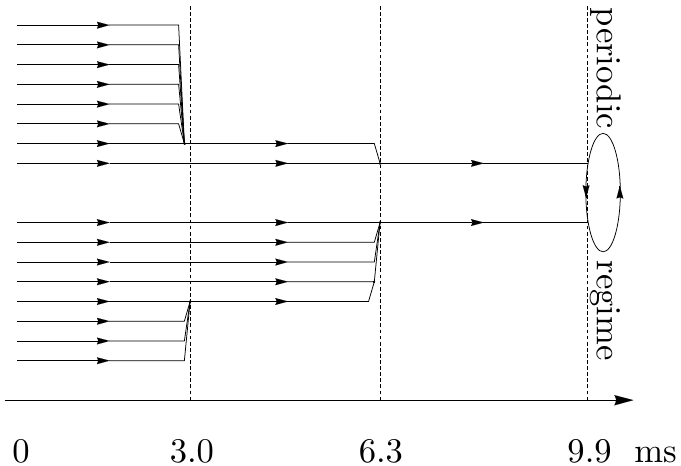}  
  \caption{\label{route} Routes to common periodic regime.
    (A) - state \#83; (B) - state \#34; (C) - state \#96, not all trajectories are shown.}
\end{figure}
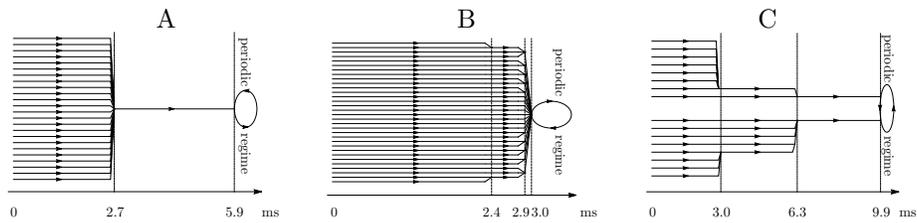

Several possible routes for the system to evolve into a periodic regime were observed.
Three examples are given in Fig. \ref{route}.
The first one is realized if all trajectories merge into a single one
well before the periodic regime is reached, see Fig. \ref{route}, (A).
The second one is realized if all trajectories merge into a smaller number of 
different
trajectories, which evolve some time before entering the cycle
at the same point, see Fig. \ref{route}, (B).
The third one is realized if all trajectories  enter the cycle
at two different points, see Fig. \ref{route}, (C).

\begin{figure}
\begin{center}
 \unitlength=1mm
  \begin{picture}(0,0)(0,0)
  \put(0,3){A}
  \end{picture}\\
  \includegraphics[width=0.7\textwidth]{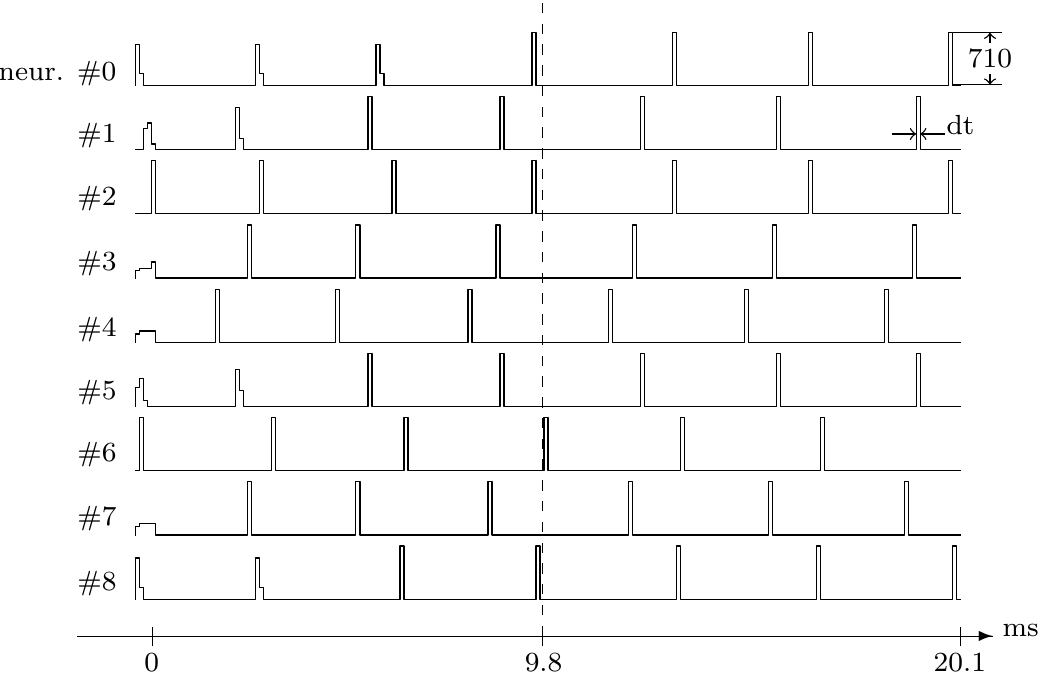}
 \begin{picture}(100,10)(0,0)
    \put(12,3){B}
    \put(50,3){C}
    \put(95,3){D}
  \end{picture}\\
  \includegraphics[width=0.3\textwidth]{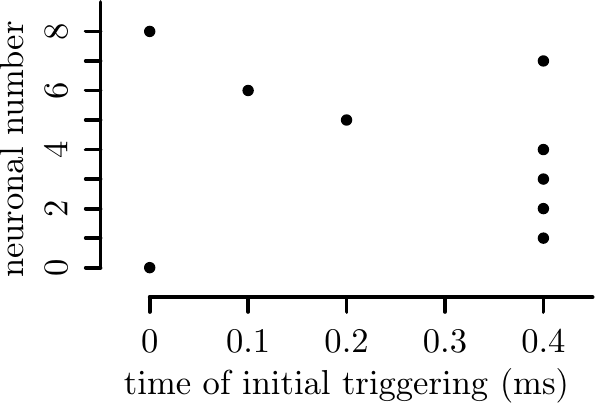}\hfill
  \includegraphics[width=0.3\textwidth]{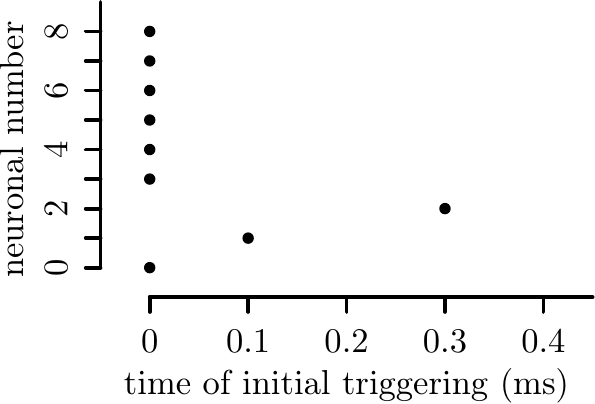}\hfill
  \includegraphics[width=0.3\textwidth]{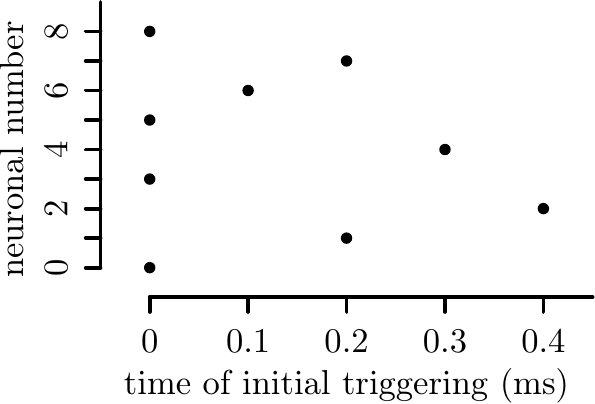}
\end{center}
  \caption{\label{disp} (A) Spiking moments of the neurons at all trajectories 
  corresponding to a single domain.
    The domain for this regime (\#6) includes 710 different stimuli.
    The periodic regime starts after 9.8 ms and has period 10.4 ms.
(B)-(D) -- examples of the initial stimuli delivering this regime.
Notice that the moment $t=0$ in (A) corresponds to the last triggering moment
in the stimulus, namely, to $t=0.4$ ms in (B), (D), and to $t=0.3$ ms in (C).
    }
\end{figure}
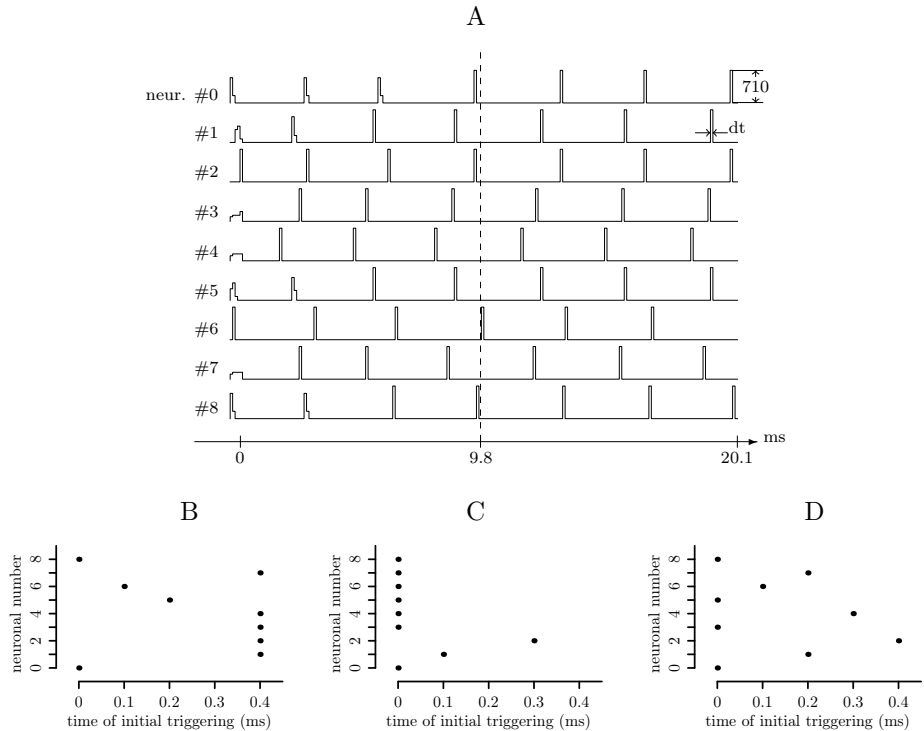

Two states were considered different in the analysis 
if the state of any axon or neural element did not match.
Therefore, synchronous firing of all corresponding neurons in two different
trajectories still does not imply that the trajectories are already merged.
The same is valid for establishing periodicity: a reproduced pattern of firing
still does not guarantee that a periodic regime is attained.
An example of the latter can be observed in Fig. \ref{disp}. 

\subsection{Triggering is necessary for merging}
 In order to test the working hypothesis, 
 all trajectories corresponding to the 102 periodic end-states were saved into separate files.
A program has been developed which analyzes what has happened just before two or more
trajectories merge into a single one. The program was applied to trajectory
sets corresponding to each periodic regime. The output of the program has shown us
that at the moment of merging, one or more neurons in the net is triggered.
Also, triggering was observed, which was not followed by merging. This leads to
a conclusion that triggering is the necessary prerequisite for merging.


\subsection{Stability}\label{stabi}
Finally, the robustness of the periodic end-states was tested.
The numerical simulation is performed in whole numbers.
Therefore we can use only finite perturbation of a state
in order to check its stability. The state of our network
is described by the values of time to live of impulses in all
axons as well as excitation level in each neuron. We have checked separately
the stability with respect to axonal and neuronal perturbations
 of all 102 periodic regimes found.
In order to perturb a single neuron's state we calculated the excitation
voltage $V$ corresponding to state variables $(n,i)$ (see Sec. \ref{intLIF} above)
and changed its value by up to $\pm 1\%$. The new $(n,i)$ pair 
was found, corresponding to the new $V$ and used as a new (perturbed) neuronal state.
After that the net was allowed to execute a free
running until a periodic state or the dead state was achieved. If the achieved
periodic regime was the same as used for perturbation, and this takes place for perturbation
of any of 9 neurons the network is composed of 
and for all states the cycle is composed of,
the unperturbed regime
was considered as stable with respect to neuronal perturbation. The
 obtained results are given in Tab. \ref{stabV}.
The smallest possible perturbation of neuronal state can be achieved if in the $(n,i)$
pair the $i$ value is replaced with $i\pm 1$. This corresponds to 
perturbation of physical voltage by 2.49$\cdot 10^{-10}\%$ to 2.51$\cdot 10^{-10}\%$.
Under perturbation of this kind, all 102 periodic states have appeared to be stable.

\begin{table}
\begin{TAB}(@,1cm,1cm)[5pt]{|c|c|c|c|}{|c|c|c|c|c|c|}
voltage perturbation size & 1\% & 0.5\% &0.1\%\\
stable & 10 & 47 & 81  \\
\parbox{3cm}{unstable, total}& 92 & 55 & 21\\
\parbox{6cm}{unstable due to approaching another known periodic regime}& 36 & 35 & 20\\
\parbox{6cm}{unstable due to approaching another unknown periodic regime}& 7 & 1 & 0\\
\parbox{6cm}{unstable due to approaching the dead state}& 49 & 19 & 1\\
\end{TAB}
\caption{\label{stabV}
Result of checking stability with respect to perturbation of neuronal voltage.}
\end{table}

In order to check stability with respect to axonal perturbation, we have
perturbed the time to live of an impulse (if any) in the axon by $\pm dt$.
If the time to live was just equal to the axonal delay, then only $-dt$ perturbation
was used. If the time to live was zero, then only $+dt$ was used.
We have proceeded further similarly as described above for neuronal states perturbation.
The result is shown in Tab. \ref{stabAx}.

\begin{table}
\begin{TAB}(@,1cm,1cm)[5pt]{|c|c|}{|c|c|c|c|c|c|}
axonal perturbation size & $dt=0.1$ ms\\
stable & 31    \\
\parbox{3cm}{unstable, total}& 71\\
\parbox{6cm}{unstable due to approaching another known periodic regime}& 43\\
\parbox{6cm}{unstable due to approaching another unknown periodic regime}& 8\\
\parbox{6cm}{unstable due to approaching the dead state}& 20\\
\end{TAB}
\caption{\label{stabAx}
Result of checking stability with respect to perturbation of axonal states.
Here, the time to live of impulse in axon was perturbed by one $dt$.}
\end{table}

We also have tried the same system with the same set of stimuli, but with
the LIF relaxation time $\tau=200$ ms instead of 20 ms. A set of 67 different
periodic end-states have been obtained with periods 3.0, 3.2, 3.4, 3.6, 4.8, 6.6. 7.2, 10.4
ms, which is rather close to what is listed in the Tab. \ref{peri}.
This might be expected since possible period values are restricted by
the values of propagation times, which were the same.

\section{Discussion and Conclusions}
Dynamics is investigated of a pulse-propagating neural net propelled
into reverberating evolution by different initial stimuli. 
Apart from final end-states in which all spiking activity was extinguished, two self-organized behaviors were observed: 1) merging of several dynamical trajectories into a single common trajectory; 2) entrainment of several initially different trajectories into a single, common periodic regime.

At the beginning of this study, it was expected
 that convergence/merging of different trajectories happens due to
neuronal firing. This idea is briefly discussed in
\cite[Sec. 2.1.1]{Vidybida2011} for the binding neuron model. 
Firing of a single neuron happens at the moment of receiving the last
(triggering) input impulse. Positions of impulses, which precede the triggering
one, may deviate slightly without changing the triggering moment, see Fig. \ref{SiCon}.
This way, two different input stimuli into a neuron are merged into a single output
spike emitted at the same moment of time. 
Due to consecutive merging in neurons the network is composed of, two different dynamics
may converge and eventually merge into a single one.
The results obtained confirm that spiking is necessary for 
the merging of trajectories in this simulation. This conclusion is made by
strict analysis of the trajectory files just before merging and after that.
In all merging cases, we have observed that some neurons have a bit different
subthreshold
excitation at different trajectories. These neurons were simultaneously
 triggered before merging and that small difference in excitation was forgotten.
For a single neuron, this mechanism is illustrated in Fig. \ref{SiCon}.

\begin{figure}[h]
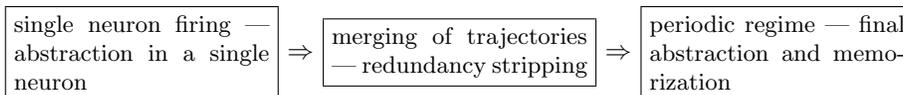

\framebox{\parbox{3.4cm}{\small single neuron firing --- abstraction in a single neuron}}
$\Rightarrow$
\framebox{\parbox{3.4cm}{\small merging of trajectories --- redundancy stripping}}
$\Rightarrow$
\framebox{\parbox{3.4cm}{\small periodic regime --- final abstraction and memorization}}
\caption{\label{RP}A possible route to perception as an emergent phenomenon due to self-organization in the time domain.}
\end{figure}

In the signal/information-processing context, neuronal firing results 
in emitting a spike, which can be treated as an abstract representation
of a set of input spikes into the neuron \cite{Vidybida2014}.
Due to cascade of such abstractions, several different trajectories can merge
into a single one, which can be treated as removing redundancy 
present in the initial stimuli. 
This same mechanism can be responsible for the noise tolerance observed 
during perception, \cite{Barlow1972,Barlow2002}.
Final attainment of a periodic regime
might be treated as an effect underpinning perception, 
see Fig. \ref{RP},
 since periodic/repetitive firing
results in formation of the short-term memory due to long-term potentiation
 \cite{Andersen2003}.
 A kind of memory is believed to be required for perception, \cite{Barlow2002}.

The simplified neuronal dynamics and signal processing assumptions 
embodied in the simulation have some obvious limitations:

1. The system relies on recurrent connections and reverberation rather than feed-forward dynamics.
While the recurrence does have a role in perception, see, e.g. \cite{Bullier2001,Bullier2001a}
for vision and \cite{Siegel1982} for hearing, the feed-forward mode of
processing is definitely present. For a discussion of the feed-forward 
mode see \cite{Abeles1982b,Aertsen1996}. Both modes can be combined, see, e.g. \cite{Zheng2014}.

2.The number of neural elements is small ($n=9$).
Actually, it is established for the IT, \cite{Gross1984},  and expected for other senses,
\cite{Barlow1972}, that objects are presented
by activity of small neuronal sets. May it be as small as 9?
A further investigation is required with a larger network.

3. Certainly, real neural networks, unlike these simulations, 
do not reproduce a periodic regime forever.
 In reality, one should expect a regime close to periodic, 
which is eventually interrupted by a mechanism not considered here.

4. The only structural change we mention here is at the final stage of processing, 
namely the long-term potentiation due to periodic regime.
It is possible that faster, spike timing dependent facilitations and inhibitions might play a role in neural dynamics even on short time scales.
A possible fast mechanism is the twitching of dendritic spines, \cite{Crick1982a}.

At the same time, we see that dynamics of a reverberating neural network can be rich
with different regimes emerging due to spatio-temporal self-organization.
This endows reverberating networks with sophisticated abilities to process signals/information.
These remain to be further investigated.\bigskip

{\small\bf Acknowledgments.} {\small AV thanks to Demiurge Technologies AG for support with attendance
of several neuroscience conferences.}





\end{document}